\documentclass[12pt]{article}
\usepackage{setspace}
\usepackage{amsmath,amssymb}
\begin{document}
\setlength{\topmargin}{-1cm} 
\setlength{\oddsidemargin}{-0.25cm}
\setlength{\evensidemargin}{0cm}

\newcommand{\e}{\epsilon}
\newcommand{\beq}{\begin{equation}}
\newcommand{\eeq}[1]{\label{#1}\end{equation}}
\newcommand{\bea}{\begin{eqnarray}}
\newcommand{\eea}[1]{\label{#1}\end{eqnarray}}
\renewcommand{\Im}{{\rm Im}\,}
\renewcommand{\Re}{{\rm Re}\,}
\newcommand{\diag}{{\rm diag} \, }
\def\draftnote#1{{\color{red} #1}}
\def\bldraft#1{{\color{blue} #1}}

\begin{titlepage}
\begin{center}

\vskip 4 cm

{\Large \bf  On the Central Charge of Spacetime Current Algebras and Correlators in String Theory on $AdS_3$  }

\vskip 1 cm

{Jihun Kim$^a$  and Massimo Porrati$^{a,b}$}

\vskip .75 cm

{$^a$ \em Center for Cosmology and Particle Physics, \\ Department of Physics, New York University, \\4 Washington Place, New York, NY 10003, USA}

\vskip .75 cm

{$^b$ \em School of Natural Sciences, Institute for Advanced Study\\ Princeton NJ USA 08540}\footnote{Member until May 2015, on sabbatical leave from NYU.}\end{center}

\vskip 1.25 cm
%\title{}
%\author{}
%\date{}Killing
%\maketitle

\begin{abstract}
\noindent  Spacetime Virasoro and affine Lie algebras for strings propagating in $AdS_3$ are known to all orders in 
$\alpha'$. The central extension of such algebras is a string vertex, whose expectation value can depend on the number of long strings
present in the background but should be otherwise state-independent. In hep-th/0106004, on the other hand, a 
state-dependent expectation value was found. Another puzzling feature of the theory is lack of cluster decomposition
property in certain connected correlators. This note shows that both problems can be removed by 
defining the free energy of the spacetime boundary conformal field theory as the Legendre transform of the formula 
proposed in the literature. This corresponds to pass from a canonical ensemble, where the number of 
fundamental strings that  create the background can fluctuate, to a microcanonical one, where it is fixed. 
\end{abstract}
\end{titlepage}
\newpage

%%%%%%%%%%%%%%%%%%%%%%%%%%%%%%%%%%%%%%%%%%%%%%%%%%%
%\section{Preliminary Results}
%%%%%%%%%%%%%%%%%%%%%%%%%%%%%%%%%%%%%%%%%%%%%%%%%%%
Bosonic strings and superstrings compactified on $AdS_3 \times M$ with  a nonzero 
Kalb-Ramond field strength on $AdS_3$ 
 allow for a world-sheet description exact in $\alpha'$. For simplicity,  we will work here with the bosonic string, but our
 argument works equally for superstrings.
The conformal field theory living on the target space $M$  is unitary while the $AdS_3$ part of the background is a Wess-Zumino-Witten $SL(2,\mathbb{R})$ model, which is exactly soluble. Many properties
of the model are known, including  its complete spectrum~\cite{mo1}, its thermal partition function~\cite{mo2} and  correlation functions~\cite{mo3}.

The spacetime affine Lie algebra and spacetime Virasoro generators
were found in~\cite{gks} in the free field Wakimoto representation~\cite{wak} 
of the $SL(2,\mathbb{R})$ WZW model. Such 
representation is adequate to study spacetime algebras, since their generators are
 non-normalizable vertices supported in
the near-boundary, weakly-coupled region of $AdS_3$. The exact form of these vertices
was found in~\cite{ks}.
In both representations, a non-vanishing central charge was found. Such central charge is exact to all orders in $\alpha'$
and exists already at tree level in the string coupling parameter, $g_S$. It is the string theory generalization of the
classical central charge found in asymptotically $AdS_3$ General Relativity by Brown and Henneaux~\cite{bh}. 

Instead of asymptotic in- or out- states, labeled by momenta and other quantum numbers, (Euclidean) 
$AdS_3$ possesses local
operators, which define a boundary CFT. They are labeled by the boundary coordinate $x\in \mathbb{C}$.  
In the string worldsheet description, local operators are  represented by vertices, also labeled by $x$. A natural set of
operators has form
 (see~\cite{ks} for notations and more details on the formalism)
\beq
V(x,\bar{x},h,I)=\int d^2 z  \Phi_h(x,\bar{x}|z,\bar{z}) O_I. 
\eeq{m1}
The operator $\Phi_h(x,\bar{x}|z,\bar{z})$ is a worldsheet 
Virasoro and $SL(2,\mathbb{R})$ affine Lie primary, with worldsheet 
conformal weight $\Delta=\bar{\Delta}=-h(h-1)/(k-2)$. Here $k$ is the level of the worldsheet $SL(2,\mathbb{R})$ 
affine Lie algebra (not to be confused with the {\em spacetime} affine Lie algebra). The operators $O_I$ belong to the conformal theory on $M$. They have conformal dimension $\Delta_I=\bar{\Delta}_I= 1+ h(h-1)/(k-2)$; since they
commute with all operators we shall use, they can be ignored in most of our discussion. The worldsheet 
$SL(2,\mathbb{R})$ currents define the operator $J(x|z)= 2xJ_3(z)-J^+(z) -x^2 J^-(z)$. The current algebra 
Operator Product Expansion (OPE) writes compactly as
\beq
J(x|z)J(y|w)= k{(y-x)^2\over (z-w)^2} + {1\over z-w} [(y-x)^2 \partial_y -2(y-x)] J(y|w) + \mbox{ regular terms}.
\eeq{m1a}

The spacetime central charge is proportional to the vertex 
\beq
I={1\over k^2}\int d^2 z J(x|z) \bar{J}(\bar{x}|\bar{z}) \Phi_1(x,\bar{x}|z,\bar{z}). 
\eeq{m2}
The $SL(2,\mathbb{R})$ current algebra level $k$ is the ratio $l^2/\alpha'$, with $l$ the radius of $AdS_3$.  
As proven in~\cite{ks}, $I$ is independent of $x,\bar{x}$. 
The $I$ vertex can be written in terms of the $\bar{\partial}$ derivative of an operator $\Lambda(x,\bar{x}|z,\bar{z})$. It is nevertheless nonzero because $\Lambda$ is not a good observable; in particular, its two-point function is logarithmic
in $z,\bar{z}$~\cite{ks}. In the near-boundary, weakly coupled region, it is conveniently written in terms of Wakimoto
variables~\cite{wak,ks} ($\beta,\gamma,\phi$) as $\lim_{\phi\rightarrow\infty}\Lambda =(x-\gamma)^{-1}$.  The operators
$\Lambda$ and $\Phi_1$  are related by~\cite{ks}
\beq
\bar{J}\Phi_1= {k \over \pi}\partial_{\bar{z}}\Lambda,
\eeq{m3}
so the identity vertex is
\beq
I={1\over k^2}\int d^2 z J(x|z) \bar{J}(\bar{x}|\bar{z}) \Phi_1(x,\bar{x}|z,\bar{z})= -{1\over 2\pi i k} \oint dz J\Lambda .
\eeq{m4}
When inserted into correlation functions of vertices~(\ref{m1}), the integral does not vanish because the operator 
product expansion of $J\Lambda$ with $\Phi_h$ has poles and because $J\Lambda$ transforms anomalously
under coordinate transformations~\cite{gk}. 

The first property follows from the OPEs~\cite{ks} ($\sim$ denotes equality up to regular terms)
\bea
\Lambda(x,\bar{x}|z,\bar{z}) \Phi_h(y,\bar{y}|w,\bar{w})&\sim &{1\over x-y} \Phi_h(y,\bar{y}|w,\bar{w}),
\nonumber \\
J(x|z) \Phi_h(y,\bar{y}|w,\bar{w})&\sim &{1\over z-w} [(y-x)^2 \partial_y +2h(y-x)] \Phi_h(y,\bar{y}|w,\bar{w}).
\eea{m5}

The second one follows because the OPE of $J(x)$ with $\Lambda(x)$ is singular~\cite{ks}
\beq
J(x|z) \Lambda(x,\bar{x}| w,\bar{w})\sim -{1\over z-w} .
\eeq{m6}
So, even though $J$ and $\Lambda$ transform under holomorphic changes of coordinates as tensors of weight
one and zero respectively\footnote{This can be seen most easily using the Wakimoto representation. Notice that
$\Lambda$ is nevertheless a bad observable because, among other things, its two-point function contains logarithmic
terms that need an IR regularization.}, the normal ordered product $:\!\! J(x|z)\Lambda(x,z)\!\!:\equiv J\Lambda(x|z)$ transforms anomalously
as
\beq
 T(z) J\Lambda(x|w) \sim  {1\over (z-w)^2} J\Lambda(x|w) + {1\over z-w} \partial_w [J\Lambda (x|w)] - {1\over (z-w)^3}.
\eeq{m7}
 So under an infinitesimal diffeomorphism $\epsilon$, $J\Lambda$ transforms as
 \beq
 \delta J\Lambda(x|z) = \partial_z \epsilon(z) J\Lambda(x|z) +  \epsilon(z) \partial_z J\Lambda(x|z) -
 {1\over 2} \partial_z^2 \epsilon(z).
 \eeq{m7a}
 Under a finite change of coordinates $z\rightarrow z'=\phi(z)$ eq.~(\ref{m7a}) integrates to
 \beq
 J\Lambda(x|z)\rightarrow (J\Lambda)'(x|z')=\partial_z \phi (z) \left[J\Lambda(x|z) +
 {1\over 2} {\partial_\phi^2 z \over \partial_\phi z}\right].
 \eeq{m7b}
    Put together, eqs.~(\ref{m5},\ref{m7b}) give rise to a puzzling result found in~\cite{gk}: the operator $I$ 
 is not proportional to the identity. We can prove this by considering the correlator
 \beq
 \langle I \prod_i \Phi_{h_i} (x_i,\bar{x}_i | z_i, \bar{z}_i) \rangle = 
 {1\over k^2}\langle \int d^2 z J(x|z) \bar{J}(\bar{x}|\bar{z}) \Phi_1(x,\bar{x}|z,\bar{z})\prod_i \Phi_{h_i} (x_i,\bar{x}_i | z_i, \bar{z}_i) \rangle .
 \eeq{m8}
 In ref.~\cite{gk} it was evaluated on a genus zero surface, but the computation can be done for arbitrary genus using the 
 Schottky parametrization of Riemann surfaces. In such parametrization, a genus $g$ surface is represented as the region
 of the complex plane outside a set of $2g$ circles $\hat{C}_n$, $n=1,..,2g$.\footnote{We take them to be all of finite radius
 so that the point $z=\infty$ lies outside all circles.} The circles are identified pairwise by 
 $SL(2,\mathbb{C})$  transformations, $z\rightarrow (az+b)/(cz+d)$, $ad-bc=1$, $c\neq 0$, 
 that map the outside of one circle in the pair to the inside of the other. 
 Using equation~(\ref{m4}), correlator~(\ref{m8}) can then be written as 
 \beq
  -{1\over 2\pi i k} \oint_C dz \langle J\Lambda (x|z)\prod_i \Phi_{h_i} (x_i,\bar{x}_i | z_i, \bar{z}_i) \rangle ,
  \eeq{m9}
  where the contour $C$  is the union of small circles $C_i$ surrounding the operator insertion points $z_i$, plus the circles
  $\hat{C}_n$, 
  {\em plus} the limit for $R\rightarrow \infty$ of a circle $C_R$ at radius $|z|=R$. The OPEs (\ref{m5}) give~\cite{gk}
  \beq
  -{1\over 2\pi i}\oint_{C_i} dz  J\Lambda (x|z)\Phi_{h_i} (x_i,\bar{x}_i | z_i, \bar{z}_i) = (h_i-1) \Phi_{h_i} (x_i,\bar{x}_i | z_i, \bar{z}_i).
  \eeq{m10}
  
To find the contribution of the circles $\hat{C}_n$ we apply eq.~(\ref{m7b}) to the the transformations that identify such 
circles pairwise. Under the map $z\rightarrow z'=(az+b)/(cz+d)$ the integrals of the homogenous term in~(\ref{m7b}) 
cancel and one is left with $g$ integral $-2^{-1}(2\pi i)^{-1}\oint_{\hat{C}'_n} dz' [2c/(-cz'+a)] $.  
Since the point $z'=a/c$ is mapped to the  point $z=\infty$, it is inside the circle $\hat{C}_n'$, so that the integral gives a contribution 
$-1$. The sign is $-$ because, under the $SL(2,\mathbb{C})$ map, the image contour runs clockwise. 
Since we have $g$ such integrals, we get
\beq
-{1\over 2\pi i}\oint_{\cup_j C_j \cup_n \hat{C}_n} dz 
\langle J\Lambda (x|z)\prod_i\Phi_{h_i} (x_i,\bar{x}_i | z_i, \bar{z}_i) \rangle = 
[\sum_j(h_j-1) -g]\langle J\Lambda (x|z)\prod_i\Phi_{h_i} (x_i,\bar{x}_i | z_i, \bar{z}_i) \rangle .
\eeq{m10a}

Finally, the integral on $C_R$ is evaluated  by performing the conformal inversion $z\rightarrow z'=-1/z$. Thanks to eq.~(\ref{m7b}), the integral becomes
\beq
\lim_{R\rightarrow \infty}{1\over 2\pi i}\oint_{C_R} dz J\Lambda (x|z) =
-\lim_{R\rightarrow \infty} {1\over 2\pi i} \oint_{|z|=1/R} dz[J\Lambda (x|z) -1/z]= 1.
\eeq{m11}
The sign here is $+$ because the integral over $C_R$ is performed clockwise. 
We thus get an all-genera version of the $g=0$  result of~\cite{gk}
\beq
\langle I \prod_i \Phi_{h_i} (x_i,\bar{x}_i | z_i, \bar{z}_i) \rangle = {1\over k} [1-g+ \sum_j(h_j-1)] \langle  \prod_i \Phi_{h_i} (x_i,\bar{x}_i | z_i, \bar{z}_i) \rangle ;
\eeq{m12}
So, the ``identity" $I$ is not constant but instead assumes 
different values on different irreducible 
representations of the Virasoro algebra.

This result is quite disastrous, because it contradicts  the fact that in the field theory limit 
$\alpha' \rightarrow 0$, the Brown-Henneaux calculation shows a unique central charge for all the irreducible
representations corresponding to light fields. Among them there are many for which one could use instead 
eq.~(\ref{m12}). More generally, a Hilbert space that decomposes into a sum of irreducible representations of the 
Virasoro algebra, each one with a different central charge, is incompatible with having a local 2D CFT on the $AdS_3$
boundary. Notice that the operator $I$ can (and does) take different values on sectors containing a 
different number of long strings~\cite{ks,gks}.

In this note, we point out that the problem has a solution. It does not involve any subtlety 
in the calculation
of ref.~\cite{gk}, though the latter assumes that the OPE of the ``bad" operator $\Lambda$  with $\Phi_h$
has no logarithmic branch cuts.\footnote{This is true in the large-$\phi$ limit, as can be checked using the Wakimoto 
representation.} Rather, the solution is that formulas for the AdS/CFT correspondence proposed in the 
literature ~\cite{dbort} do not specify how to treat the operator $I$. A standard generalization would treat the identity $I$ as any other vertex, add a source for it, and define a ``free energy." We will show that instead the correct generalization is a Legendre transform of the free energy.   This change corresponds to pass from a canonical ensemble, where the number of 
fundamental strings that produce the background can fluctuate, to a microcanonical one, where it is held fixed. 
Our proposal also resolves another puzzle 
of the standard definition, namely the lack of cluster decomposition in some connected correlators.

To begin with, recall that the spacetime correlators contain contributions from disconnected worldsheet topologies.
In fact, the holographic correspondence proposed in~\cite{dbort} states that the generating functional for the vacuum correlators of the spacetime CFT, $Z$, is the {\em exponential} of the string 
partition function\footnote{An easy way to see this is to notice that the string partition function contains only one integration over the zero modes of spacetime coordinate fields per each connected component of the worldsheet.}
\bea
W &=&  \sum_{g=0}^\infty g_S^{2g-2}\langle \exp\left[\int d^2x J(x,\bar{x},h,I) V(x,\bar{x},h,I) + \int d^2x \lambda(x) I\right] \rangle_g,\nonumber \\
Z&=&C\exp(W).
\eea{m13}
The expectation value $\langle ... \rangle$ is computed by performing the functional integral of the worldsheet action
over {\em connected} closed Riemann surfaces of genus $g$; $g_S$ is the string coupling constant; 
the constant $C$ is arbitrary. We recognize in $W$ the generator of connected correlators for the spacetime CFT, that is
the free energy. Even though we introduced a local source $\lambda(x)$ for the vertex $I$, the free energy depends 
only on $\lambda_0\equiv \int d^2x \lambda(x)$, since $I$  is independent of $x,\bar{x}$. 

Next, consider the correlators 
\bea
\langle\langle \prod_{i=1}^N \int d^2 z_i \Phi_{h_i} (x_i,\bar{x}_i | z_i, \bar{z}_i) O_I \rangle\rangle
&\equiv& \prod_{i=1}^N\left. {\delta \over \delta J(x_i,\bar{x}_,h_i,I_i)} Z \right|_{J=0} , \label{m13a} \\
\langle\langle I\prod_{i=1}^N \int d^2 z_i \Phi_{h_i} (x_i,\bar{x}_i | z_i, \bar{z}_i) O_I \rangle\rangle
&\equiv& \prod_{i=1}^N\left. {\delta \over \delta J(x_i,\bar{x}_i,h_i,I_i)} {\delta \over \delta\lambda(x,\bar{x})}
Z \right|_{J,\lambda=0}, \label{m13b} \\
\langle\langle 1 \rangle \rangle &\equiv& \left. Z\right|_{J=\lambda=0}.
\eea{m14}
For simplicity, assume that the disconnected components of correlator~(\ref{m13a}) vanish.
Then, the expectation value~(\ref{m13b})
is the sum of two pieces~\cite{dbort}
\bea
&&\langle\langle I\prod_{i=1}^N \int d^2 z_i \Phi_{h_i} (x_i,\bar{x}_i | z_i, \bar{z}_i) O_I \rangle\rangle=
 \langle I \rangle \langle\langle \prod_{i=1}^N \int d^2 z_i \Phi_{h_i} (x_i,\bar{x}_i | z_i, \bar{z}_i) O_I \rangle\rangle + \nonumber \\
&& \langle \langle 1 \rangle \rangle \sum_g{1\over k}[1-g+\sum_i(h_i-1)]g_S^{2g-2}\langle \prod_{i=1}^N \int d^2 z_i \Phi_{h_i} (x_i,\bar{x}_i | z_i, \bar{z}_i) O_I \rangle_g.
\eea{m15}
The expectation value over connected components includes here a sum over genera so that e.g. 
$dW/d\lambda_0 \equiv \langle I \rangle \equiv \sum_{g=0}^\infty g_S^{2g-2} \langle I \rangle_g $  and the leading term 
in $\langle I \rangle$ is ${\cal O}(g_S^{-2})$.
Since the last term  in eq.~(\ref{m15}) comes from connected VEVs, it would be absent if $I$ were truly proportional to the identity. In~\cite{ks} it was argued that a connected component  $I(g_S)\langle \prod_{i=1}^N \int d^2 z_i \Phi_{h_i} (x_i,\bar{x}_i | z_i, \bar{z}_i) O_I \rangle_g$ is permissible, as long as $I(g_S)$ is the same for all correlators containing at least
an insertion of either $I$ or $\Phi_h$, but this is not compatible 
with $I$ being the identity as the following argument shows.
Consider the correlator $\langle \langle I^n \rangle \rangle$ 
for arbitrary integer $n\geq 1$. If $I$ were the identity then
$\langle \langle I^n \rangle \rangle=A B^n$, for some constants $A$ and $B$. 
On the other hand, from the definition of connected correlators we have $W =D \exp[\lambda_0 I(g_S)]$, with $D$
a constant; thus we arrive at the recursion relation
\beq
\langle \langle I^{n} \rangle \rangle= \langle \langle 1\rangle \rangle I(g_S)^n \left. P_n(x)\right|_{x=D}, \qquad
\langle \langle I^{n+1} \rangle \rangle=\langle \langle 1\rangle \rangle I(g_S)^{n +1}
\left[ \left. xP_n(x) + x{d\over dx}P_n(x)\right] \right|_{x=D} ,
\eeq{m15a}
with $P_n[x]$ a polynomial of degree $n$ in $x$ such that $P_0=1$. By computing the VEVs for $n=1,2,3$ we immediately 
find that they cannot be of the form $AB^n$. 

The fact that $I$ has nonzero {\em connected} correlators with physical vertices has another troubling consequence.
In fact, as shown in~\cite{mo3}, the four point function of operators with weight $h_1,..,h_4$ in the spacetime CFT factorizes (when $h_i+h_j <(k+1)/2$) on operators belonging to the discrete series $1/2<h<(k-1)/2$, as well as on other operators. Among the former is the operator $I$, so the four point function factorization is 
\beq
W_{1234} (x_1,x_2,x_3,x_4)= W_{12\lambda} (x_1,x_2){1\over W_{\lambda\lambda}}W_{34\lambda}(x_3,x_4) +
1\leftrightarrow 3+1\leftrightarrow 4+ ....
\eeq{fact}
Here $....$ means a sum over other factorization channels and 
we used the shorthand $W_i= \delta W/\delta J(x_i,\bar{x}_i,h_i)$, 
$W_\lambda ={\delta W / \delta \lambda}(x)$ etc.~\footnote{Eq.~(\ref{fact}) was derived in~\cite{mo3} to lowest order in 
the $g_S$ expansion, so to compare our results with known formulas one must truncate them and keep 
the ${\cal O}(g_S^2)$ term only. }
Since $W_{ij\lambda}(x_i,x_j,x) $ and $W_{\lambda\lambda}(x)$ 
are independent of $x$, the {\em connected} correlator $W_{1234}$ does not obey the cluster decomposition property. 

So we must kill in a fell swoop all connected correlators containing the operator $I$. To see how to achieve this,  
 we must recall first that spacetime Virasoro and affine-Lie algebra currents are also represented by vertices, 
 $T_{xx}(x)$ and $K_{a\, x}(x)$, whose explicit form is given in~\cite{ks}. The sources for these vertices are: 
$g_{\bar{x}\bar{x}}(x,\bar{x})$,
 transforming as the boundary 2D metric, for the Virasoro vertex and $A^a_{\bar{x}}(x,\bar{x})$, transforming as a 2D
 gauge field, for the affine-Lie algebra vertex. 
 
 We will deal here with the affine Lie algebra Ward identity; the Virasoro Ward identity can be treated in a similar
 fashion.
 
 By denoting with  $\delta_\epsilon A_{\bar{x}}= D_{\bar{x}}^a \epsilon_a $ the gauge variation of the source 
 $A_{\bar{x}}$ and with
 $\delta_\epsilon J^I $ the variation of the sources of vertices~(\ref{m1}), we can write the Ward identity for the free energy
 $W$ as
 \beq
 {\cal G}_\epsilon W[A] = 0, 
  \qquad
 {\cal G}_\epsilon = \int d^2x \delta_\epsilon A(x,\bar{x}){\delta \over \delta A(x,\bar{x}) }+ 
 \delta_\epsilon J^I (x,\bar{x}){\delta \over \delta J^I(x,\bar{x})}. 
 \eeq{m15b} 
 Actually, this equation is wrong because the spacetime current algebra contains a central term, which is reflected in the vertex identity~\cite{ks}  
 \beq
 \langle ...K^{a}(x)K^b(y)... \rangle = \langle ... {1\over (x-y)^2} k_G I  + 
 {1\over x-y} f^{ab}_{\;\;\;c} K^c(y)... \rangle .
 \eeq{m16}
 It generates an anomaly in the conservation law of the current sourced by $A_{\bar{x}}$~\cite{fad}.
 So, the definition of ${\cal  G}_\epsilon$ must be modified as follows: 
 we make $\lambda(x)$, the source of the ``identity'' vertex $I$, change under gauge transformations as
 \beq
 \delta_\epsilon \lambda(x,\bar{x}) =  -\pi k_G \epsilon_a(x) \partial_{x} A^a_{\bar{x}}(x). 
 \eeq{m16a}
 Thus, the Ward identity generator  ${\cal G}_\epsilon$ changes into 
 \beq
 {\cal G}_\epsilon\rightarrow {\cal G}'_\epsilon = {\cal G}_\epsilon + \int d^2 x  
 \delta_\epsilon \lambda(x,\bar{x}) {\delta \over \delta \lambda(x,\bar{x})}.
 \eeq{m16b}
 Because of its transformation law~(\ref{m16a}), $\lambda$ is a Green-Schwarz~\cite{gs} field, which cancels the anomaly; therefore, the Ward identity is 
 ${\cal G}'_\epsilon W = 0$. In fact, an anomalous term in the Ward identity would be  
 ${\cal G}'_\epsilon W = \Delta(\epsilon)$, with $\Delta(\epsilon) $ a local functional of the 
 background gauge field $A_{\bar{x}}$ only,  which obeys the standard Wess-Zumino consistency conditions~\cite{wz}. Such term
 is canceled by adding to $W$ a term linear in $\lambda$. 
 
  The free energy $W$ obeys another identity: thanks to eqs.~(\ref{m12},\ref{m13}), we have
\beq
{\delta W\over \delta \lambda(x,\bar{x} )}=-{1\over 2k}g_S{\partial W \over \partial g_S}+ 
\sum_{h,I}{(h-1) \over k}\int d^2y J(y,\bar{y},h,I){\delta W\over \delta J(y,\bar{y}, h,I)} .
 \eeq{m17}
The solution to this linear equation is
\beq
W[\lambda, g_S, A(x), J(x,\bar{x}, h,I)] = 
W\left[0, e^{-\lambda_0/2k}g_S, A(x), e^{(h-1)\lambda_0/k}J(x,\bar{x}, h,I)\right].
\eeq{m18}
If $I$ were a central term, the generating functional would obey ${\cal G}_\epsilon W= K \delta \lambda$, with $K$
the (numerical) coefficient of the gauge anomaly. Instead we have 
\beq
{\cal G}_\epsilon W = -\int d^2 x  
 \delta_\epsilon \lambda(x,\bar{x}) {\delta W \over \delta \lambda(x,\bar{x})}=-\int d^2 x  
 \delta_\epsilon \lambda(x,\bar{x}) {dW\over d\lambda_0}.
 \eeq{m19}
 
 The observables we are interested in are the correlators of the vertices~(\ref{m1}); the source $\lambda$ is just a convenient trick to write a simple Ward identity. In fact, an object at least as natural as $W(\lambda,J)$ is a functional
 that depends on the VEV of $I$ instead of $\lambda$: the Legendre transform of $W$, that we call the ``effective action"
 \beq
 \Gamma[ \langle I \rangle , J]= W[\lambda_0, J] - \lambda_0 \langle I \rangle, \mbox{ computed at }
 {dW\over d \lambda_0}= \langle I \rangle.
 \eeq{m20} 
 Now the Ward identity on $\Gamma$ has the correct form
 \beq
 {\cal G}_\epsilon \Gamma = -\int d^2x \delta \lambda (x) \langle I \rangle  .
 \eeq{m21}
 
 The VEV $\langle I \rangle$ is essentially the total number of fundamental strings creating the $AdS_3$ 
 background. At tree level each additional long string state adds +1 to the VEV while a short string state adds a 
 "fraction of a long string"  equal to $(h-1)/k$~\cite{gk}.  Legendre transforming in $\lambda$ corresponds to defining 
 $\Gamma$ in a microcanonical ensemble where the string number is fixed. The free energy $W[\lambda]$ is instead 
 defined in a  canonical ensemble where such number can fluctuate while the ``chemical potential'' $\lambda$ is held fixed.
 Clearly, we can expect a standard CFT only when the central charge (proportional to the number of 
 fundamental strings) is fixed, not when it fluctuates. Related issues were 
 discussed in the context of precision counting of black hole microstates in~\cite{sen}.~\footnote{We thank J. Maldacena 
 for this remark and for bringing to our attention reference~\cite{sen}.}
 Given the similarity between the operator $I$ in the Wakimoto representation and the area operator of Liouville (see e.g.
 eq. (3.1) in ref.~\cite{gk}), our
 definition is analogous to defining Liouville theory at fixed area.~\footnote{This analogy was pointed out to us by D. Kutasov.}  
 
 Besides  the anomaly equation, the connected correlators of vertices~(\ref{m1})  also change, 
 because they are now defined by varying $\Gamma$ with  respect to the sources $J$ at fixed $\langle I \rangle$. 
 Using the definition of the Legendre transform~(\ref{m20}), the same shorthand notation as before 
 and the fact that the spacetime CFT has 
 vanishing one-point functions, we can expand $W-\lambda \langle I \rangle$  around $J=0$, $\lambda=0$ as
 \beq
 \Gamma= W[0,J] + \sum_{ij}{1\over 2} W_{\lambda ij} [0,0]J^i J^j \lambda + 
 {1\over 2}W_{\lambda\lambda}[0,0]\lambda^2 + {\cal O}(J^2\lambda^2, J^3\lambda) , 
 \mbox{ computed at } {dW \over d\lambda} = \langle I \rangle. 
 \eeq{exp}
  This formula shows that the two and three-point correlators of vertices~(\ref{m1}) are unchanged. 
  The four-point function changes as
  \beq
  \Gamma_{1234} (x_1,x_2,x_3,x_4)= W_{1234}(x_1,x_2,x_3,x_4) - W_{12\lambda}(x_1,x_2) {1\over W_{\lambda\lambda} }W_{34\lambda} (x_3,x_4)-  { 1\leftrightarrow 3 - 1\leftrightarrow 4} .
  \eeq{4point}
  Comparing with eq.~(\ref{fact}) we see that the non-clustering term cancels out. This cancelation holds in general. 
  In fact, by construction $W$ generates tree level connected correlators of $I$, so that its Legendre transform $\Gamma$ 
  generates 1PI irreducible correlators, containing no internal lines for the field $I$.
  This is an important check of our proposal: it not only solves the ``identity problem'' but also takes care of the breakdown
  of cluster property in the spacetime CFT. Of course these problems are related, they both originate from the fact that $I$
  has non-vanishing connected correlators with physical vertices. 
 
 So, finally we can write a partition function that obeys all standard properties of a spacetime CFT living on the boundary
 of $AdS_3$ as
 \beq
 Z= C \exp \left(\Gamma[\langle  I \rangle , J]\right).
 \eeq{m23}
 
 We conclude with two comments on this formula. 
 
 The first one is that one must compute the effective action $\Gamma[\Phi,J]$  at $\Phi = \langle I \rangle$. Computing 
 $\Gamma[\Phi,J]$ at 
 $\Phi \neq \langle I \rangle$ results in an unphysical theory, with the wrong value for the anomaly and without 
 cluster decomposition property. In fact the cancelation between dangerous
 terms in eqs.~(\ref{fact},\ref{4point}) holds only for $\Phi = \langle I \rangle$. 
 
 The second is that our prescription is valid for $k>1$. For $k<1$, the identity is not a physical operator so it does 
 not appear in eq.~(\ref{fact}); therefore, the right definition for the boundary spacetime CFT may not involve a Legendre
 transform after all. This fact may play a role in explaining some of the unusual properties of strings on $AdS_3$ at
 $k<1$~\cite{gkrs}.
 \section*{Acknowledgements}
We thank A. Giveon, D. Kutasov,J. Maldacena, N. Seiberg and E. Witten for useful comments and suggestions. 
J-H.K. is supported by an NYU JAGA fellowship. 
M.P. is supported in part by NSF grant PHY-1316452.

%%%%%%%%%%%%%%%%%%%%%%%%%%%%%%%%%%%%%%%%%%%%%%%%%%%

\end{document}